%% file: main.tex
\documentclass[usenatbib]{mnras}

\usepackage{graphicx}	
\usepackage{amsmath}	
\usepackage{amssymb}	
\usepackage{bm}		
\usepackage{xcolor}
\usepackage{soul}
\usepackage{threeparttable}
\usepackage[normalem]{ulem}
\def\ahf{\texttt{AHF}}

\def\fire{\texttt{FIRE}}
\defcitealias{Hopkins:2017}{H17}
\newcommand{\kms}{{\rm km\,s^{-1}}}
\newcommand{\rh}{\mathrm{r_{1/2}}}

\newcommand{\lcdm}{$\Lambda$CDM}
\newcommand{\msun}{{\rm M}_{\odot}}

\newcommand{\mvir}{M_{\rm vir}}
\newcommand{\rvir}{R_{\rm vir}}
\newcommand{\mstar}{M_{\star}}
\newcommand{\vmax}{V_{\rm max}}

\newcommand{\hopkins}{\citetalias{Hopkins:2017}}

\usepackage[T1]{fontenc}
\usepackage{ae,aecompl}

\usepackage{txfonts}

\title[No Assembly Required: Dwarf Galaxy Merger Histories]{No Assembly Required: Mergers are Mostly Irrelevant for the Growth of Low-mass Dwarf Galaxies}

\author[A. Fitts et al.]{Alex Fitts$^1$\thanks{\href{mailto:afitts@astro.as.utexas.edu}{afitts@astro.as.utexas.edu}}, 
Michael Boylan-Kolchin$^1$\thanks{\href{mailto:mbk@astro.as.utexas.edu}{mbk@astro.as.utexas.edu}}, James S. Bullock$^2$, Daniel R. Weisz$^{3}$, \newauthor
Kareem El-Badry$^3$, Coral Wheeler$^4$, Claude-Andr\'e Faucher-Gigu\`ere$^5$,\newauthor Eliot Quataert$^3$, Philip F. Hopkins$^4$, Du\v{s}an Kere\v{s}$^6$, Andrew Wetzel$^{7}$, Chris Hayward$^{8,9}$\\
$^1$Department of Astronomy, The University of Texas at Austin, 2515 Speedway, Stop C1400, Austin, Texas 78712-1205, USA\\
$^2$Department of Physics and Astronomy, Center for Cosmology, 4129 Reines Hall, University of California Irvine, CA 92697, USA\\
$^{3}$Department of Astronomy, 501 Campbell Hall, University of California, Berkeley, CA, 94720, USA\\
$^4$TAPIR, California Institute of Technology, Pasadena, CA, USA\\
$^5$Department of Physics and Astronomy and CIERA, Northwestern University, Evanston, IL, USA\\
$^6$Department of Physics, Center for Astrophysics and Space Sciences, University of California, San Diego, La Jolla, CA, USA\\
$^7$Department of Physics, University of California, Davis, CA, USA\\
$^8$Center for Computational Astrophysics, Flatiron Institute, 162 Fifth Avenue, New York, NY 10010, USA\\
$^9$Harvard-Smithsonian Center for Astrophysics, 60 Garden Street, Cambridge, MA 02138, USA
}
\date{\today}
\vspace{-0.5cm}
\pubyear{2018}

\begin{document}
\setstcolor{red}
\label{firstpage}
\pagerange{\pageref{firstpage}--\pageref{lastpage}}
\maketitle

\begin{abstract}
We investigate the merger histories of isolated dwarf galaxies based on a suite of 15 high-resolution cosmological zoom-in simulations, all with masses of $M_{\rm halo} \approx 10^{10}\,\msun$ (and M$_\star\sim10^5-10^7\,\msun$) at $z=0$, from the Feedback in Realistic Environments (\fire) project. The stellar populations of these dwarf galaxies at $z=0$ are formed essentially entirely "in situ": over 90$\%$ of the stellar mass is formed in the main progenitor in all but two cases, and all 15 of the galaxies have >70$\%$ of their stellar mass formed in situ. Virtually all galaxy mergers occur prior to $z\sim3$, meaning that accreted stellar populations are ancient. On average, our simulated dwarfs undergo 5 galaxy mergers in their lifetimes, with typical pre-merger galaxy mass ratios that are less than 1:10.  This merger frequency is generally comparable to what has been found in dissipationless simulations when coupled with abundance matching. Two of the simulated dwarfs have a luminous satellite companion at $z=0$. These ultra-faint dwarfs lie at or below current detectability thresholds but are intriguing targets for next-generation facilities. The small contribution of accreted stars make it extremely difficult to discern the effects of mergers in the vast majority of dwarfs either photometrically or using resolved-star color-magnitude diagrams (CMDs). The important implication for near-field cosmology is that star formation histories of comparably massive galaxies derived from resolved CMDs should trace the build-up of stellar mass in one main system across cosmic time as opposed to reflecting the contributions of many individual star formation histories of merged dwarfs.

\end{abstract}


\section{Introduction}
\input{table1.tex}
The resolved stellar populations of Local Group dwarfs provide a plethora of information related to their origin and evolution. Observations provide an `archaeological' study of their antecedents and have informed our understanding of faint galaxies at early times \citep{Hodge:1989, Bullock:2000, Freeman:2002, Ricotti:2005, Madau:2008, Tolstoy:2009,Bovill:2011, Brown:2012,Benitez-Llambay:2015}.  In recent years, it has become increasingly apparent that the study of these nearby galaxies as windows into the high-redshift universe, also known as near-field cosmology, will provide complementary opportunities to direct observations in the next generation of high-redshift galaxy surveys \citep[e.g.,][]{Weisz:2014c,Boylan-Kolchin:2015,Patej:2015,Boylan-Kolchin:2016,Graus:2016}. This translates to the near field being one of the most interesting frontiers when it comes to questions of reionization and high-redshift galaxy formation. Yet any attempt to address these questions relies directly on accurately dissecting the star formation histories (SFHs) of nearby dwarf galaxies. For this to be effective, it is necessary to understand the underlying origin of these SFHs and, specifically, whether they can be treated as individual, rather than composite, populations.  

In the \lcdm~paradigm, galaxies are the result of baryons condensing in the very center of potential wells formed by dark matter halos \citep{White:1978,Blumenthal:1984}. As a result, a galaxy's mass assembly is heavily influenced by the underlying host halo's mass assembly. The early phase of halo assembly is characterized by rapid halo growth dominated by major mergers, while the late phase is characterized by slower quiescent growth predominantly through accretion of material onto the outer portions of the halo \citep{Zhao:2003}. This buildup of mass proceeds in a hierarchical fashion; smaller structures form first and merge to form increasingly more massive structures. Accordingly, structure formation is largely self-similar across all mass scales \citep[e.g.,][]{Stewart:2008,Wetzel:2009,Fakhouri:2010}.

Simulations of massive galaxies reveal that galaxy assembly also follows a similar `two-phase' formation: first, galaxies undergo a phase of intense, in situ star formation, which is then followed by a phase of accretion of old, less massive and therefore more metal-poor systems \citep{Naab:2007,Oser:2010,Lackner:2012}. However, galaxy formation is decidedly not self-similar across mass scales,  which can be seen in the explicit nonlinear mapping of galaxy mass to halo mass \citep{Purcell:2007,Behroozi:2013b,Moster:2013}. Hence, the details of galaxy assembly cannot be determined from the halo assembly alone and can differ greatly depending on the mass scale. 

Dwarf galaxies themselves differ a great deal from their more massive counterparts: they are far more dark matter dominated, are much fainter (10-10$^7$ times fainter than Milky Way (MW)-mass galaxies) and are inefficient at forming stars from their large gas reservoirs (e.g., \citealt{Hunter:1985,Blanton:2001,Skillman:2003,Robertson:2008}). It is therefore unclear whether the same assembly processes observed in massive galaxies scale down to the dwarf regime. Already, a range of studies have found that the accretion of `fresh' gas from the IGM is highly dependent on halo mass: while MW-mass halos can efficiently accrete baryons at late times (e.g., \citealt{Keres:2005,Dekel:2006,vandeVoort:2011,Faucher-Giguere:2011,Wetzel:2015a,AnglesAlcazar:2016}), gas accretion becomes increasingly inefficient at late times in halos below $\sim10^{11}\,\msun$ \citep{Bullock:2000,Hoeft:2006,Noh:2014,El-Badry:2018}.  Mergers, too, are thought to have an increasingly small effect on galaxy assembly given the sharp decline in the $M_\star-M_\mathrm{halo}$ relation at low galaxy masses \citep{Hopkins:2008,Stewart:2009,Brook:2014}.

At the same time, however, mergers may explain a number of observed features in dwarf galaxies. Interactions specifically between dwarfs have been suggested to explain observed gas bridges and shells \citep{Besla:2012,Pearson:2016} and are also thought to restart star formation in certain `two-component' dwarfs \citep{Benitez-Llambay:2015}.  Even interactions with dark halos have been predicted to increase the star formation rate \citep{Starkenburg:2016a} and leave observable asymmetries \citep{Starkenburg:2016b}. Galaxy mergers are also thought to be one explanation for older, metal poor stars being in the outskirts of dwarf galaxies (\citealt{Benitez-Llambay:2016}, though see \citealt{El-Badry:2016}). It is therefore important to have a concrete understanding of the merger histories of dwarf galaxies in order to explore whether they are a viable candidate to explain such features.

While galaxy mass assembly in simulations often is analyzed in terms of the contributions from in situ star formation and accreted stellar growth, this has been explored much less in dwarfs as compared to massive galaxies. Previous attempts to constrain the galaxy merger history of nearby dwarfs have either relied on abundance matching with dark-matter-only (DMO) simulations \citep{Behroozi:2013b,Moster:2013,Deason:2014} or semi-analytical models \citep{Hirschmann:2013}. These abundance matching models have pointed to a diminished importance of accreted stellar mass in isolated low-mass galaxies, but with the exact slope and scatter of the M$_\star$-M$_\mathrm{halo}$ relation at the low-mass end still hotly debated \citep{Behroozi:2013b,Brook:2014,Garrison-Kimmel:2017a,Munshi:2017,Moster:2017}, lingering questions remain.

Sidestepping the present M$_\star$-M$_\mathrm{halo}$ debate, hydrodynamical simulations have aimed to simulate the physical properties of isolated dwarf galaxies from first principles. Some simulations (e.g., \citealt{Simpson:2013,Jeon:2017}) have found that dwarf galaxies are built up from a diverse patchwork of smaller merging galaxies. However, \citet{AnglesAlcazar:2016}'s simulation of a dwarf spheroidal, involving a particle tracking analysis of a set of FIRE simulations, have produced supporting evidence for galaxy mergers playing a minor role in dwarf galaxy assembly.  Given the wide array of dwarf properties possible at a fixed halo mass \citep{Onorbe:2015,Fitts:2017}, it is both useful and necessary to study a larger sample of dwarf galaxies in order to understand both the spread in galaxy merger histories possible for dwarf galaxies and how it ultimately affects their stellar populations.
\begin{figure*}
\includegraphics[width=0.99\textwidth]{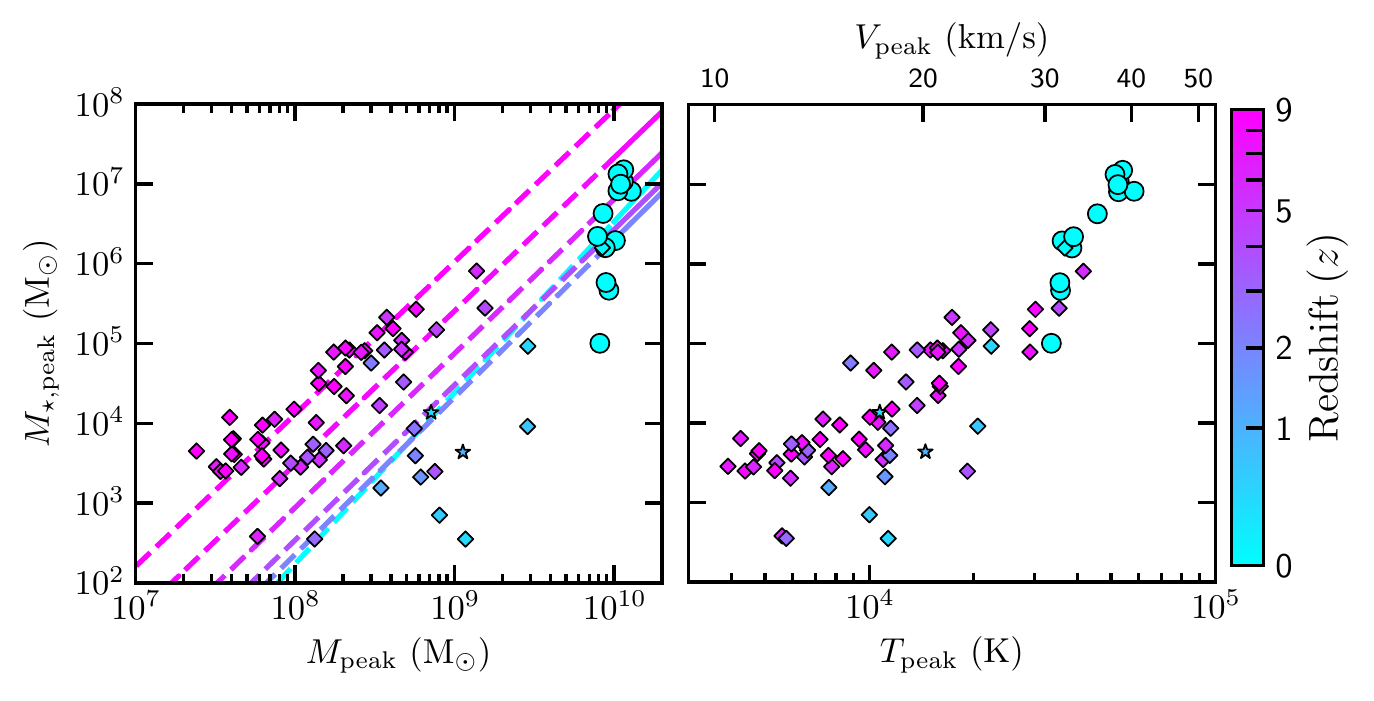}
  \caption{\textit{Left}: M$_\mathrm{\star,peak}$-M$_\mathrm{peak}$ relation for all of the galaxies in the merger trees of the 15 central halos in our simulated suite of dwarfs. The 15 central dwarf galaxies are plotted as circles at their $z=0$ masses. All merger progenitors are plotted as diamonds at their peak halo masses. The two star symbols represent the two satellites that still exist at $z=0$. The points are colored by their accretion time onto the main progenitor. Also included are the Behroozi et al.~(in prep) M$_\mathrm{\star}$-M$_\mathrm{halo}$ relations for $z=0,2,4,6,8,10$ (solid color lines), while dashed lines show power law extrapolations to lower masses. These lines are colored according to their redshift. Comparing to this simple extrapolation, the progenitors qualitatively follow the relation and its evolution with time, though the simulated dwarfs may populate a slightly steeper M$_\mathrm{\star,peak}$-M$_\mathrm{peak}$ relation at low redshift. \textit{Right}: The same plot but now in terms of the virial temperature of the halo at the time of peak halo mass, T$_\mathrm{peak}\equiv$T$_\mathrm{vir}(t_\mathrm{peak})$. Much of spread visible in the left panel is merely a reflection of M$_\mathrm{peak}$'s dependence on redshift. T$_\mathrm{peak}$ provides a measure that is free of pseudo-evolution. We see a strong, redshift-independent relation between T$_\mathrm{peak}$ and M$_\star$. The corresponding V$_\mathrm{peak}$, the maximum circular velocity of the dark matter halo at the time of peak halo mass, is labeled on the top x-axis.}
  \label{fig:mstar_mhalo}
\end{figure*} 

Our suite \citep{Fitts:2017} focuses solely on isolated dwarf galaxies. The field provides a pristine measure of low-mass dark matter halos, as it is free from the usual environmental factors that can influence those halos that stray into the virial radius of larger halos (e.g., tidal stripping, \citealt{Fillingham:2015}, and decreased number counts, \citealt{Garrison-Kimmel:2017b}). Various observed properties of a dwarf galaxy, such as gas content (e.g., \citealt{Grcevich:2009}), star formation history (e.g., \citealt{Grebel:2003, Weisz:2011}), and morphology (e.g., \citealt{Lisker:2007,Pearson:2016}), are also strongly correlated with proximity to a massive galaxy. By focusing on only isolated dwarfs, any possible observable consequences of mergers in our suite can be determined unambiguously.   Finally, since isolated dwarfs are twice as likely to have had a major merger as compared to satellites of similar mass \citep{Deason:2014}, our study focuses on the instances where mergers are likely to have the largest impact on galaxy assembly.

Our simulation suite makes use of the \fire-2\footnote{\url{http://fire.northwestern.edu}}~hydrodynamical simulations of galaxy formation with detailed stellar feedback implementation. \fire~\citep{Hopkins:2014} cosmological simulations of dwarf galaxies have reproduced several key observables, including realistic galactic outflows \citep{Muratov:2015,Muratov:2017}, the dense HI content of galaxy halos \citep{Faucher-Giguere:2015},  the mass-metallicity relation \citep{Ma:2016}, the mass-size relation and age/metallicity gradients \citep{El-Badry:2016}, cored dark-matter profiles \citep{Onorbe:2015,Chan:2015}, stellar kinematics \citep{Wheeler:2017}, the Kennicutt-Schmidt relation \citep{Orr:2017}, observed abundance distributions \citep{Escala:2017} and a realistic population of satellites around MW-mass hosts \cite{Wetzel:2016}

The paper is arranged as follows. Section \ref{sec:simulations} provides a brief review of our simulation suite. Section \ref{sec:results} outlines the main results of our study, including the dwarf-dwarf merger histories for our suite, along with a dedicated look at the stellar mass formed in situ separately from the stellar mass delivered from mergers. We also comment on the presence of satellites located around a number of our dwarfs. Finally, we compare to several recent works and provide a broader interpretation of our results in Section \ref{sec:discussion}.

\section{Simulations}
\label{sec:simulations}
\begin{figure}
\includegraphics[width=0.49\textwidth]{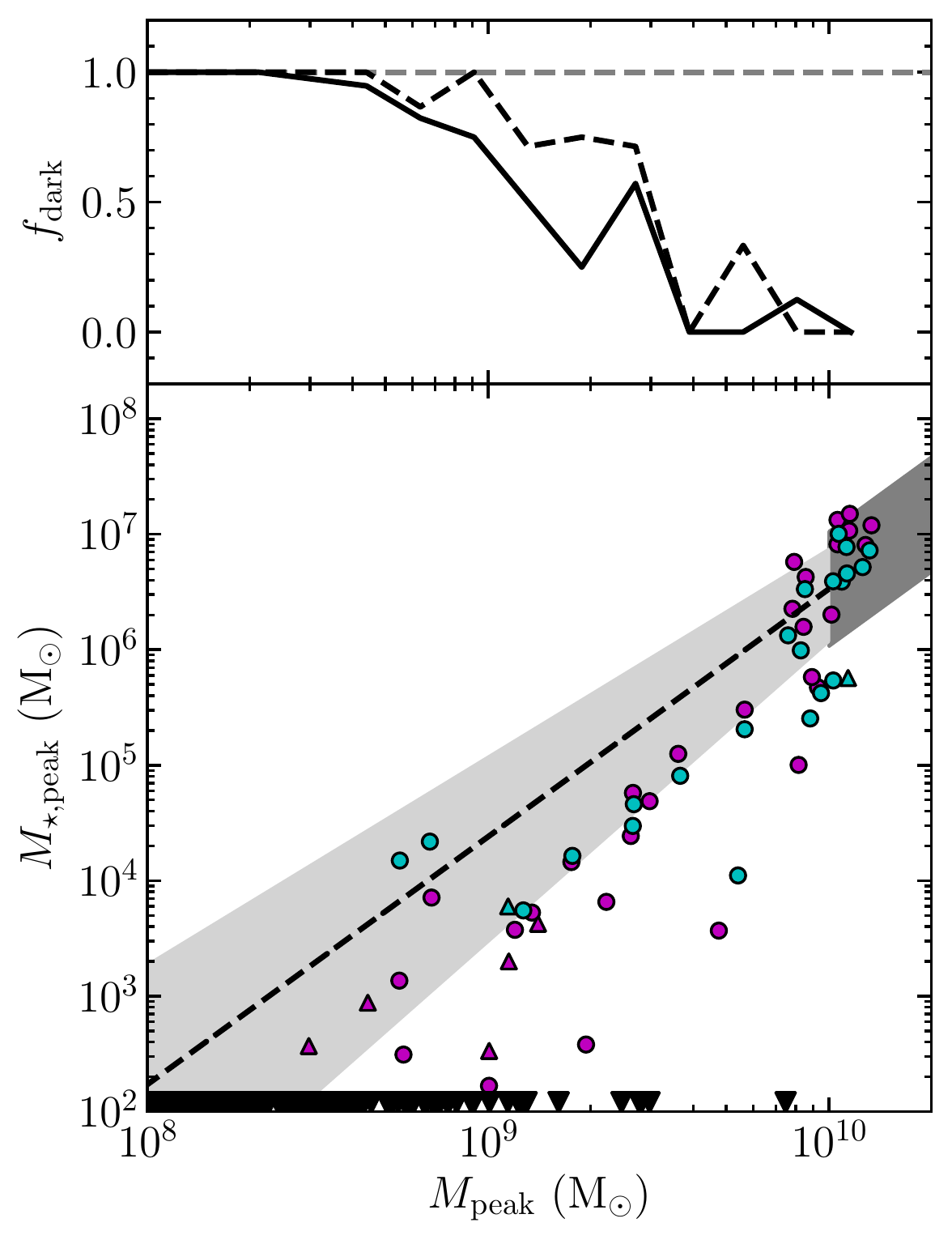}
  \caption{\textit{Bottom}: $M_{\star,\,\rm peak}-M_\mathrm{peak}$ relation at $z=0$ for all halos that have >95$\%$ high resolution particles. Magenta circles represent central galaxies while magenta triangles represent satellites, both at our fiducial resolution. The corresponding cyan symbols indicate the low-res versions of our simulations. Dark halos (from fiducial runs) are plotted as downward black triangles at the bottom of the plot. The M$_\star$-M$_\mathrm{halo}$ relation from Behroozi et al.~(in prep.) is plotted with 0.5 dex scatter in dark grey. A simple power law extrapolation is shown by the dashed black line. The light grey region depicts the possible range of slopes of the low-mass end of the relation according to \citet{Garrison-Kimmel:2017a}. Results from both resolution levels of our simulations populate a relation that appears to have additional scatter to lower stellar masses; the importance of resolution is also evident, as this scatter is more pronounced in the runs that are able to resolve lower-mass galaxies. \textit{Top}: The solid (dashed) black line gives the overall fraction of halos that do not contain any star particles and are dark at each halo mass in our fiducial (low-resolution) simulations. We find that $>95\%$ of halos with M$_\mathrm{peak}<7\times10^8\,\msun$ are dark. }
  \label{fig:fracofdark}
\end{figure} 
Our simulation suite consists of 15 cosmological zoom-in simulations of \lcdm\ dwarf galaxy halos chosen to have virial masses of $10^{10}\,\msun \:(\pm 30\%)$ at $z=0$ (see \citealt{Fitts:2017} for details).The simulations here are part of the Feedback In Realistic Environments (FIRE)\footnote{\url{http://fire.northwestern.edu}}, specifically the ``FIRE-2'' version of the code; all details of the methods are described in \hopkins, Section~2. The simulations use the code GIZMO \citep{Hopkins:2015},\footnote{\url{http://www.tapir.caltech.edu/~phopkins/Site/GIZMO.html}}, with hydrodynamics solved using the mesh-free Lagrangian Godunov ``MFM'' method.  The simulations include cooling and heating from a meta-galactic background and local stellar sources from $T\sim10-10^{10}\,$K; star formation in locally self-gravitating, dense, self-shielding molecular, Jeans-unstable gas; and stellar feedback from OB \&\ AGB mass-loss, SNe Ia \&\ II, and multi-wavelength photo-heating and radiation pressure; with inputs taken directly from stellar evolution models. The FIRE physics, source code, and all numerical parameters are {\em exactly} identical to those in \hopkins. Our fiducial simulations with galaxy formation physics included have baryonic (dark matter) particle masses of $500\,\msun$ ($2500\,\msun$), with physical baryonic (dark matter) force resolution of $h_{b}=2\,$pc ($\epsilon_{\rm DM}=35\,$pc); force softening for gas uses the fully-conservative adaptive algorithm from \citet{Price:2007}, meaning that the gravitational force assumes the identical mass distribution as the hydrodynamic equations (resulting in identical hydrodynamic and gravitational resolution). 

To ensure that we explore the physics of star formation and internal feedback separately from environmental effects, each target halo is required to be separated from any more massive halo by at least 3 times the virial radius of the more massive halo (while any more massive halo is required to lie beyond 5 times the virial radius of the target halo). The halos span a representative range of concentrations (and therefore, formation times; e.g., \citealt{Navarro:1997,Wechsler:2002}) for their mass. Initial conditions are generated with \texttt{MUSIC} \citep{Hahn:2011}, using the approach outlined in \cite{Onorbe:2014}. 

In post-processing, we identify halos and construct merger trees with the Amiga Halo Finder (\ahf; \citealt{Knollmann:2009}). By constructing merger trees, we are able to correlate the \ahf\ halo catalogs across time, allowing us to track the dwarf's mass assembly history (MAH) as well as its star formation history. We assign the primary progenitor at each snapshot by ranking all halos who contributed particles from the previous snapshot according to the following metric,
\begin{equation}
M_{ij}=\frac{N^2_{i\cap j}}{N_iN_j}
\label{eq:ahf}
\end{equation}
where $N_i$ is the number of particles in a halo, $N_j$ is the number of particles in a progenitor of that halo in the previous snapshot and $N_{i\cap j}$ is the number of shared particles. The halo that maximizes this metric function is identified as the main progenitor of the current halo in the previous snapshot. This process is repeated for the entirety of the simulation.
For each halo (subhalo), we compute the maximum (peak) halo mass ever reached by the main branch of a progenitor, M$_\mathrm{peak}\equiv$M$_\mathrm{vir}(t_\mathrm{peak})$. For the rest of this paper, we will focus on quantities that occur at the time of peak halo mass, $t_\mathrm{peak}$. We choose to focus on this time instead of the time of accretion because tidal forces can have substantial effects on merger companions even prior to the time of accretion.  In general, the tidal force from the central disk potential can strip off a portion of the outer mass of a subhalo, shifting it to lower $M_\mathrm{vir}$, or it can completely destroy the subhalo, either through tidal shocking \citep{Gnedin:1999} or repeated stripping events (though see \citealt{VandenBosch:2017} for a discussion of the difficulties associated with assessing whether simulations suffer from numerical over-merging).

In our analysis, we only consider merging halos that have a (dark matter) mass ratio of at least 1:100 with the main progenitor halo of the central dwarf at the time of peak mass. To verify that these halos are resolved, we also require the mass to be equal to or greater than the corresponding mass from the mean mass assembly history \citep{Fakhouri:2010} for a M$_\mathrm{vir}(z=0)\sim10^8$ $\msun$ halo.  Merging companions of this mass are actually quite well-resolved in dark matter, with $\gtrsim 10^{4}$ particles. This also means that $\gtrsim 10^{4}$ baryonic particles have {\em participated} in the formation history of the halo and have ``cycled through it'' (assuming something like the universal baryon fraction is associated with the halo). \hopkins\ shows that convergence in the stellar mass functions for FIRE simulations is actually good down to only a few stellar particles for dwarfs.  Specifically, at our choice in mass resolution, we can expect $\sim10^8$ $\msun$ halos at $z=0$ to be at least within a factor of $\sim3$ of the converged stellar mass. Given our interest solely in their existence, rather than detailed properties, we will include all visible merging companions that would have attained M$_\mathrm{vir}>10^8$ $\msun$ by $z=0$ and have at least 10 star particles. 
\begin{figure*}
\includegraphics[width=0.95\textwidth]{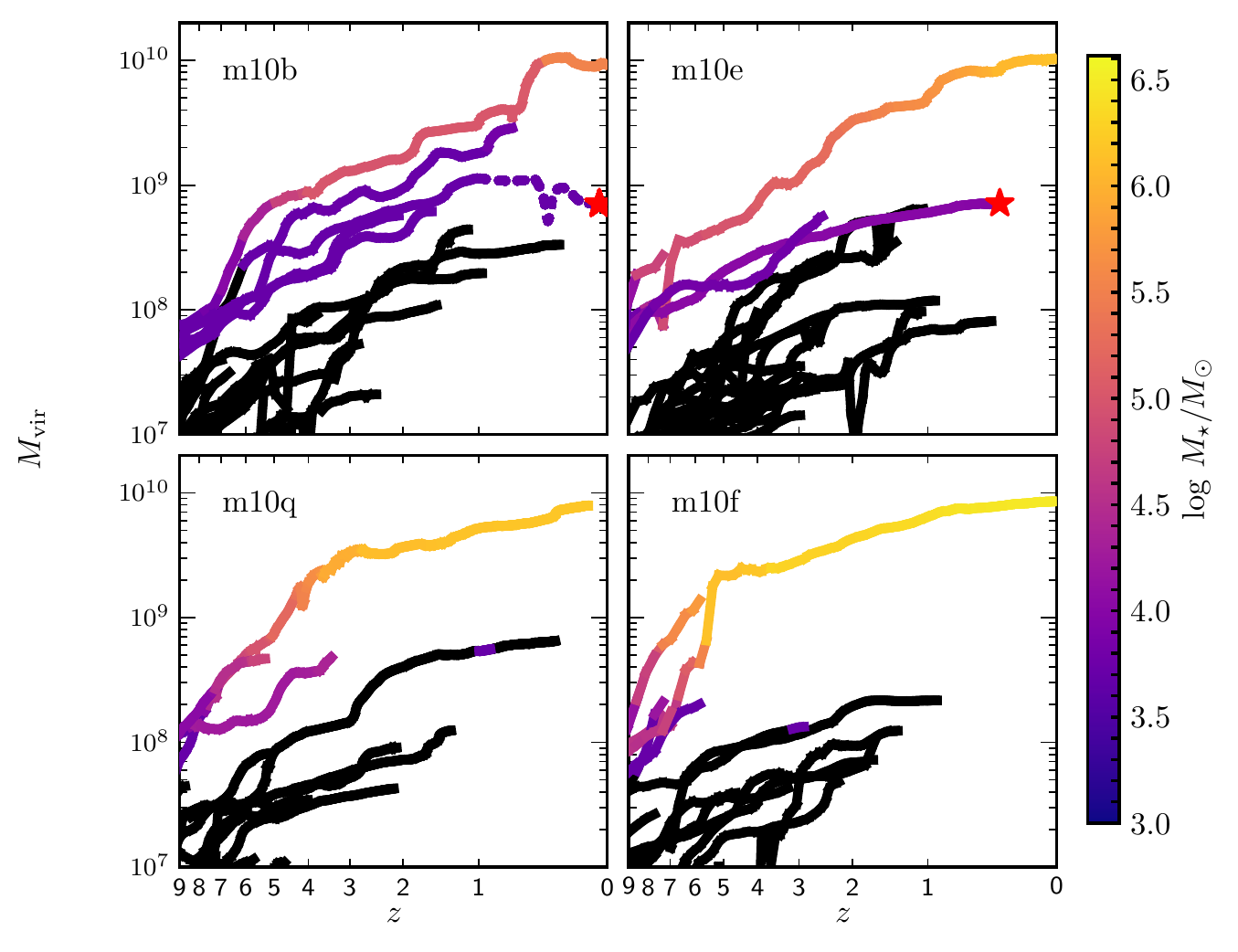}
  \caption{Mass assembly histories of the main progenitor and merging companions with at least a 1:100 halo mass ratio with the main progenitor for four example halos from our suite (clockwise from top left, in order of increasing $M_\star$: m10b, m10e, m10q, m10f). Merger companion histories are plotted up until the time of their peak mass. Histories are colored according to the instantaneous stellar mass of each galaxy. Black lines represent starless ("dark") halos; red stars correspond to luminous satellites at $z=0$. The dotted line in the upper left plot marks the second approach of m10b's present day satellite. The two upper panels display the MAHs of the two dwarfs with a companion at $z=0$, while the two lower panels show the two dwarfs that contain the most accreted stars.  Though our suite of dwarfs display a wide range of merger histories, the end result is the same in each case: the stellar mass contributed by mergers or accretion is  minimal.}
  \label{fig:mvirs_v_t}
\end{figure*} 
For convergence testing, we have run one of our halos, m10b, at 2 times poorer (better) force and 8 times poorer (better) mass resolution; these simulations are named m10b$\_$low (m10b$\_$high). The baryonic (dark matter) particle masses of $63\,M_{\sun}$ ($313\,M_{\sun}$) of m10b$\_$high makes it is one of the best-resolved cosmological simulations of a dwarf galaxy at present. We include a number of its central properties in Table~\ref{table:cprops} along with the properties of the lower resolution runs. m10b$\_$high looks very similar to the fiducial version in most respects, though it contains nearly a factor of $\sim2$ more stars at $z=0$. While at first this might appear odd, it must be noted that multiple iterations of the same initial conditions can typically yield factors of $\sim2$ difference in stellar mass at this mass scale due to the purely stochastic run-to-run variation of star formation \citep{Su:2017}. We also see a  mild decreasing trend in the stellar half-mass radius as the resolution is increased. The convergence of physical size of halo m10b is discussed further in appendix \ref{sec:appendixa}.
\section{Results}
\label{sec:results}
\subsection{The $M_\star-M_{\rm halo}$ Relation}
The galaxy-halo connection, as encapsulated by the $M_{\star,\,\rm peak}-M_{\rm peak}$ (left) and $M_{\star,\,\rm peak}-T_{\rm peak}$ (right) relationships (where $T_{\rm peak}$ is the virial temperature of the halo at $t=t_{\rm peak}$), is plotted in Figure~\ref{fig:mstar_mhalo}. In both panels, we include all galaxies that are part of the merger trees of all the 15 main dwarfs in our simulation suite. The $z=0$ dwarfs are plotted as circles while all progenitors at earlier epochs are plotted as diamonds and are colored according to their accretion time (defined as the time at which the merging companion entered the virial radius of the main progenitor). The star symbols mark the two luminous satellites that are within their host's virial radius at the present day. For reference, we have also plotted the $M_\star-M_\mathrm{halo}$ relations from Behroozi et al.~(in preparation) for $z=0,2,4,6,8$ and 10 in the left panel. The Behroozi line is constrained by observations only for $M_\star>10^{10}\,\msun$; for comparison purposes, this relation is extrapolated to lower masses (below $10^{10}\,\msun$) with a simple power law. The progenitors qualitatively follow the expected relation and its evolution with time, though the simulated dwarfs appear to populate a slightly steeper relation at low redshift.

The redshift dependence of the stellar mass-halo mass relation in the left panel of \ref{fig:mstar_mhalo} is relatively strong. However, it is well-known that M$_\mathrm{vir}(z)$ of a halo is subject to `pseudo-evolution': the reference density (in our calculations, $\rho_\mathrm{crit}$) entering the mass definition depends on redshift, and therefore, the mass of a halo will change even if there is no physical accretion. For the majority of low-mass halos (M$_\mathrm{vir}\lesssim10^{12}\,\msun$), pseudo-evolution accounts for almost all of the evolution in mass since $z = 1$ \citep{Diemer:2013,Wetzel:2015a}. To remove this pseudo-evolution behavior from our relation, we instead plot the maximum circular velocity at the time of peak halo mass (V$_\mathrm{peak}\equiv$V$_\mathrm{max}(t_\mathrm{peak})$). This quantity is similar to $z_\mathrm{vmax}$ from \citet{Li:2007} and provides a measure of the central gravitational potential, which is established substantially earlier than the final virial mass of the halo \citep{Fitts:2017}. The right panel of Fig.~\ref{fig:mstar_mhalo} exhibits a significantly tighter, redshift-independent galaxy-halo relationship. As found in \citet{Fitts:2017}, the stellar content in these halos is directly connected to their central density, specifically during the fast accretion phase when the majority of mass is aggregated.  

Fig.~\ref{fig:fracofdark} presents a closer look at the halo-galaxy relation in terms of M$_\mathrm{peak}$. It shows the M$_\mathrm{\star,\,peak}$-M$_\mathrm{peak}$ relation for all of the halos in our simulation suite that have at most 5\% contamination (by mass) from lower-resolution particles at $z=0$. Central galaxies in our fiducial simulations are colored magenta and shown as circles while satellites are plotted as triangles. Cyan symbols signify the corresponding galaxies in our low-resolution runs. Black downward triangles represent dark (starless) halos in our fiducial simulations that satisfy our resolution criteria. The M$_\star$-M$_\mathrm{halo}$ relation from Behroozi et al.~(in prep.) is plotted with 0.5 dex scatter in dark grey, while the black dashed line represents a simple power-law extrapolation. The light grey region indicates the range of slopes expected for low mass galaxies depending on the inherent scatter present in the relation (as determined by the ELVIS N-body simulations coupled with Local Group galaxy counts, see discussion in \citealt{Garrison-Kimmel:2017a}). The upper bound of the region correlates with a scatter of 0 while the lower region correlates with a scatter of 2 dex. While our results generally agree with the lower bound of the region, the scatter of the relation at the low mass end has already been shown to be dependent on resolution as well as environment \citep{Munshi:2017}. This comparison highlights the importance resolution may play in pinning down a well-defined slope for the M$_\star$-M$_\mathrm{halo}$ relation as well (see Appendix \ref{sec:appendixa} for further discussion on resolution and convergence). 

The top panel of Fig.~\ref{fig:fracofdark} displays the overall fraction of resolved halos at $z=0$ that are dark at each halo mass in our fiducial (solid) and low resolution (dashed) simulations. We find that below M$_\mathrm{vir}\sim7\times10^8$, nearly all resolved halos in our suite are dark. Similar to \citet{Sawala:2016a}, we find that the transition from luminous to dark satellites occurs at roughly M$_\mathrm{vir}\sim3\times10^9$ at $z=0$. However this result may be relatively sensitive to the timing of reionization (Elbert et al., in preparation), with earlier (later) reionization times translating to larger (smaller) transition masses, and may prove useful in constraining the specific timing of reionization by comparing to observation \citep[e.g.,][]{Tollerud:2017}. We also note that whether or not a halo is truly devoid of stars will depend sensitively on resolution and the implementation of a variety of baryonic processes. Given our convergence tests, and those in \hopkins, our definition of "dark" should be thought of as halos containing no more than $\sim 4000\,\msun$ of stars.
\subsection{Mergers}
\begin{figure}
\includegraphics[width=0.49\textwidth]{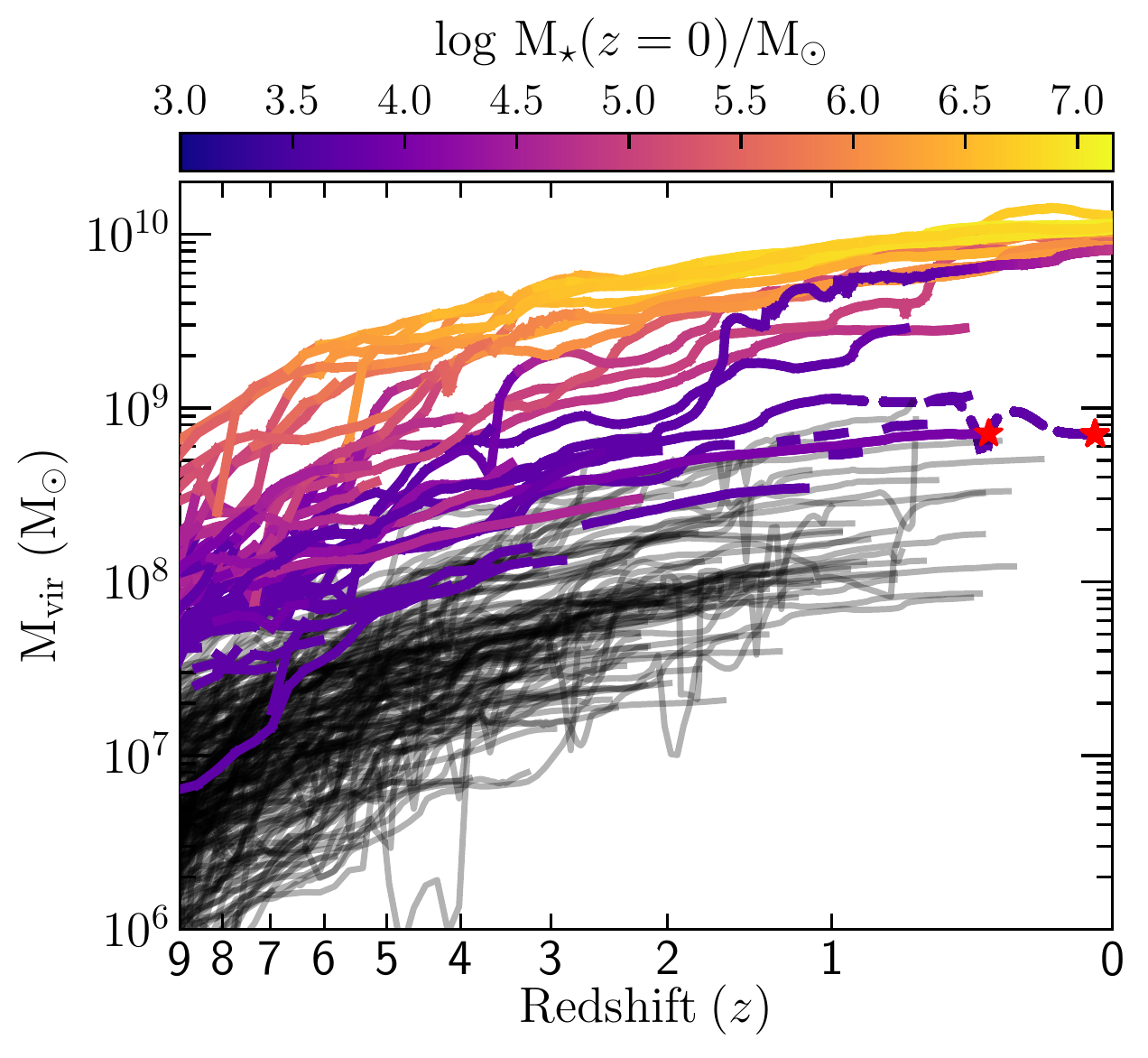}
  \caption{The mass assembly histories for all 15 isolated dwarf galaxies and their merger companions plotted until they reach their peak halo mass. Histories are colored according to their instantaneous stellar mass. Grey lines represent dark halos. Red stars represent luminous satellites at $z=0$. A cutoff in galaxy formation is apparent below a virial mass of $\sim4\times10^7\,\msun$ by a redshift of 9, suggesting that reionization  prevents gas cooling and star formation in low-mass halos.}
  \label{fig:allmvir}
\end{figure} 
\begin{figure*}

\includegraphics[width=0.995\textwidth]{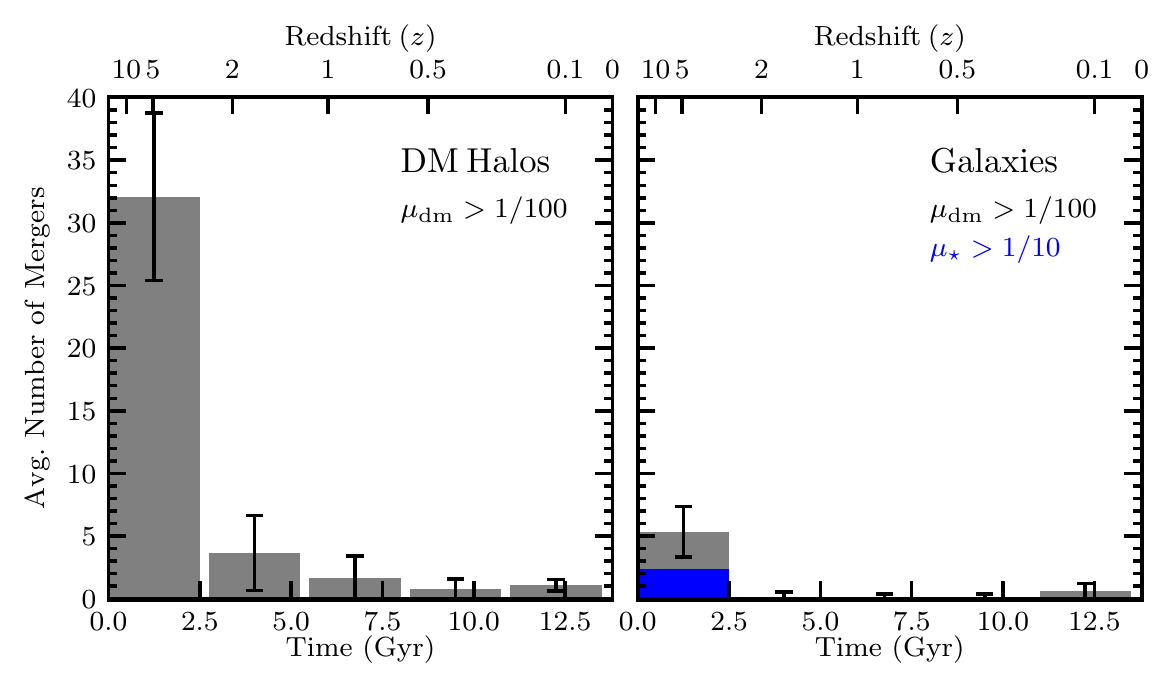}
\caption{\textit{Left}: The average number of resolved halo mergers for our suite of isolated dwarf galaxies. Black error bars show the $1\,\sigma$ spread over the suite. Only mergers with halo mass ratio greater than 1:100 are included in our count. Mergers are most common at early times, reflecting the early assembly epoch of dwarf dark matter halos. \textit{Right}: The average number of galaxy mergers that satisfy the same halo mass ratio criterion and have at least 10 star particles (in each progenitor) are plotted in grey. On average, isolated dwarf galaxies undergo few galaxy mergers, with most mergers occurring in the very earliest phase of mass assembly (and at early cosmic times). Those galaxy mergers that also have a galaxy ratio of at least 1:10 with the main progenitor are plotted in blue. It is apparent that our dwarfs experience few galaxy mergers and that the vast majority of these mergers contribute minimally to the total stellar mass.}
  \label{fig:merger_tot}
\end{figure*} 
\begin{figure}
\includegraphics[width=\columnwidth]{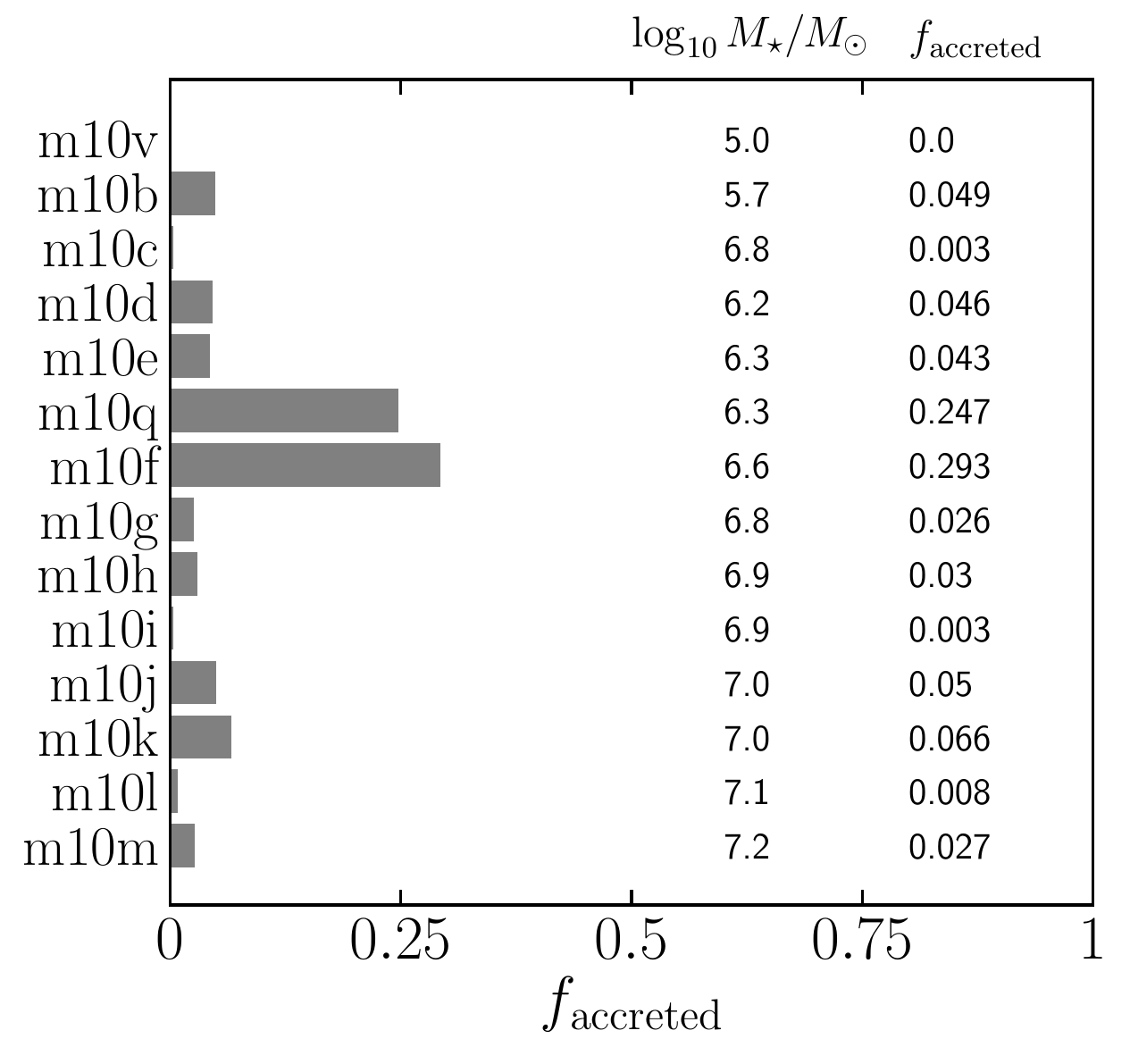}
  \caption{The fraction of $z=0$ stellar mass in our simulated dwarfs contributed by accretion events (i.e., the fraction of stellar mass formed outside of the each dwarf's main progenitor branch). Accretion only accounts for more than 7\% of $z=0$ stellar mass in two out of the 14 dwarfs that host galaxies at $z=0$. Both the stellar mass at $z=0$ and fraction of this mass coming from accretion are listed on the plot as well.}
  \label{fig:bar_insitu}
\end{figure} 
The MAHs of four of our simulated dwarfs are shown in Fig. \ref{fig:mvirs_v_t}. Accompanying each main progenitor are the MAHs of all progenitors having at least a 1:100 halo mass ratio by the time of peak mass. Each MAH is colored according to the total stellar mass within the central galaxy [defined as M$_\star(<0.1\times \rm R_\mathrm{vir})$] at each time; halos without stars are colored in black. Progenitors are plotted until the time of peak mass; histories belonging to $z=0$ satellites galaxies terminate in a red star. In the upper-left panel, the MAH of the $z=0$ satellite is plotted as a dashed line from its first approach ($z \approx 1$) till its time of infall a second and final time. The dwarfs clearly exhibit a range of merger histories. m10b, which forms one of the lower amounts of $z=0$ stellar mass (upper left panel, M$_\star\sim5\times10^5\,\msun$), has several galaxy mergers occurring after $z\sim~2$, though they contribute little to the overall stellar mass. Meanwhile, halo m10f (lower right) experiences the largest visible merger ($\sim3:1$) in our entire suite (and, accordingly, has the largest contribution from accreted stars, $\sim30\%$), but subsequently has a relatively quiescent assembly. 

If we compile the MAHs of all the progenitors that merge with each main dwarf (Fig.~\ref{fig:allmvir}), we find that a rough cutoff for galaxy formation exists. The majority of merger companions that contain a galaxy (colored lines in Fig. \ref{fig:allmvir}) all attained a virial mass of $\sim4\times10^7\,\msun$ (V$_\mathrm{max}\sim10~$km/s) by a redshift of 9. This can be understood physically given that hydrogen reionization is complete in our simulations by $z\sim~10$ (December 2011 update of the \citet{Faucher-Giguere:2009} UVB model). Any halos with a virial temperature at or below the cosmic reionization threshold of $\sim10^4$ K are unable to accrete fresh gas \citep{Bullock:2000,Hoeft:2006,Faucher-Giguere:2011,Noh:2014}. Halos with no star formation prior to reionization remain dark. Meanwhile, the constant color of the lowest-mass galaxies highlights that they experience an initial burst of star formation at early times but form few stars subsequently. These progenitors were just above the reionization suppression threshold at early times, but their SFHs are sharply truncated by the loss of baryons induced by the combined effects of reionization and feedback \citep{Benitez-Llambay:2015} and very much resemble `fossils' of reionization \citep{Ricotti:2005}. 

By compiling and binning every merger event in time, we are able to sketch the mean merger history for isolated dwarfs presented in our suite (Fig.~\ref{fig:merger_tot}). The grey bars in the left panel illustrate the average number of halo mergers for each dwarf's main progenitor over all of cosmic history.  Only mergers that have a mass ratio of at least 1:100 (defined in terms of peak mass) are considered in our analysis. The black error bars indicate one standard deviation. The majority of halo mergers occur before $z\sim2$ and clearly trace the rapid assembly phase \citep{Wechsler:2002} of these halos in this mass range. Meanwhile, the right panel of Fig. \ref{fig:merger_tot} displays all the galaxy mergers that occur with the main progenitor that also satisfy our resolution criterion of at least 10 star particles. Each of our dwarfs experiences approximately 5 galaxy mergers in its lifetime, though the vast majority of these mergers contribute very little fractional mass. 

Comparing the two panels of Fig. \ref{fig:merger_tot}, we find that dark halo mergers are nearly an order of magnitude more common than luminous galaxy mergers and occur over a longer period of time for our simulated field dwarfs. At low redshifts ($z \la 2$), galaxy mergers are extremely rare. The dearth of low-$z$ galaxy mergers is a result of the early assembly of dwarf halos combined with the majority of late-time merging occurring with halos that were insufficiently massive to form stars before reionization. These low-mass halos were therefore too small to form stars at early times but also have no opportunity to gather the material needed to form stars after reionization. If we focus on those mergers that also satisfy a \textit{galaxy} merger ratio of 1:10 (blue bars in the right panel of Fig. \ref{fig:merger_tot}), we see that the majority are minor galaxy mergers contributing few stars (fractionally). Nearly all of these events occur before redshift 3 and hence the vast majority of the stellar mass that originates outside of the main progenitor is accreted at early times during the rapid assembly phase (which is broadly consistent with what is found in the literature: \citealt{Klimentowski:2010,Deason:2014}). 

Fig. \ref{fig:bar_insitu} lists the $z=0$ fraction of stars accreted from mergers for each of our 15 dwarfs, as well as the stellar mass of each central galaxy at $z=0$.  Across our entire suite, which samples nearly 2 decades in stellar mass, there appears to be little correlation between the total number of galaxy or halo mergers and the stellar mass of the main dwarf. The great majority of our dwarfs -- 13 of 15 -- contain less than 7$\%$ of their stellar mass in stars from galaxy mergers. And even the two dwarfs that contain a higher percentage are dependent upon our exact definition of what constitutes a main progenitor. Looking at the bottom panels of Fig. \ref{fig:mvirs_v_t}, we see that halo m10f obtains most of its accreted stars from interacting with a larger halo at early times and that m10q experiences a similar major merger early in its formation that contributes a significant portion of the accreted stars. This highlights an underlying ambiguity in how to distinguish the main progenitor of our halos. For example, if we were to follow the most massive progenitor at each snapshot instead of using Eq. \ref{eq:ahf} we would find that these two halos, like the rest of our suite, would have $<10\%$ of their stellar mass from outside sources. The fractions quoted in Fig. \ref{fig:bar_insitu} therefore serve as an upper limit to the stellar contribution from mergers, in all likelihood. 

Our central finding -- that the vast majority of stellar mass is formed in the main progenitor halo -- is robust to uncertainties in merger tree definitions. It follows that the stellar mass in dwarfs of this mass is not an amalgam of populations from multiple progenitors but rather is a clean representation of star formation in one main halo across time. When attempting to infer the star formation history of an observed field dwarf galaxy with a stellar mass comparable to those of the simulations presented here, we should therefore confidently be able to attribute the overwhelming majority of its stellar population to star formation in one progenitor and not to the conditions within multiple merging companions. This conclusion is not dependent on radius within the galaxies: the fraction of stars originating from mergers never rises above 0.3 at any point within $5\,r_{1/2}$ (see also Graus et al., in preparation, for a more detailed look of how radial distance from the center can affect CMDs derived from observations). While \citet{El-Badry:2016} found that older stars tend to migrate to the outer regions of low-mass galaxies as a consequence of stellar feedback (though they did not consider galaxies with high merger rates), their results are strongest for galaxies that are more massive than those studied here.

\subsection{Stellar Populations}
\label{sec:StellarPop}
\begin{figure*}
\includegraphics[width=0.995\textwidth]{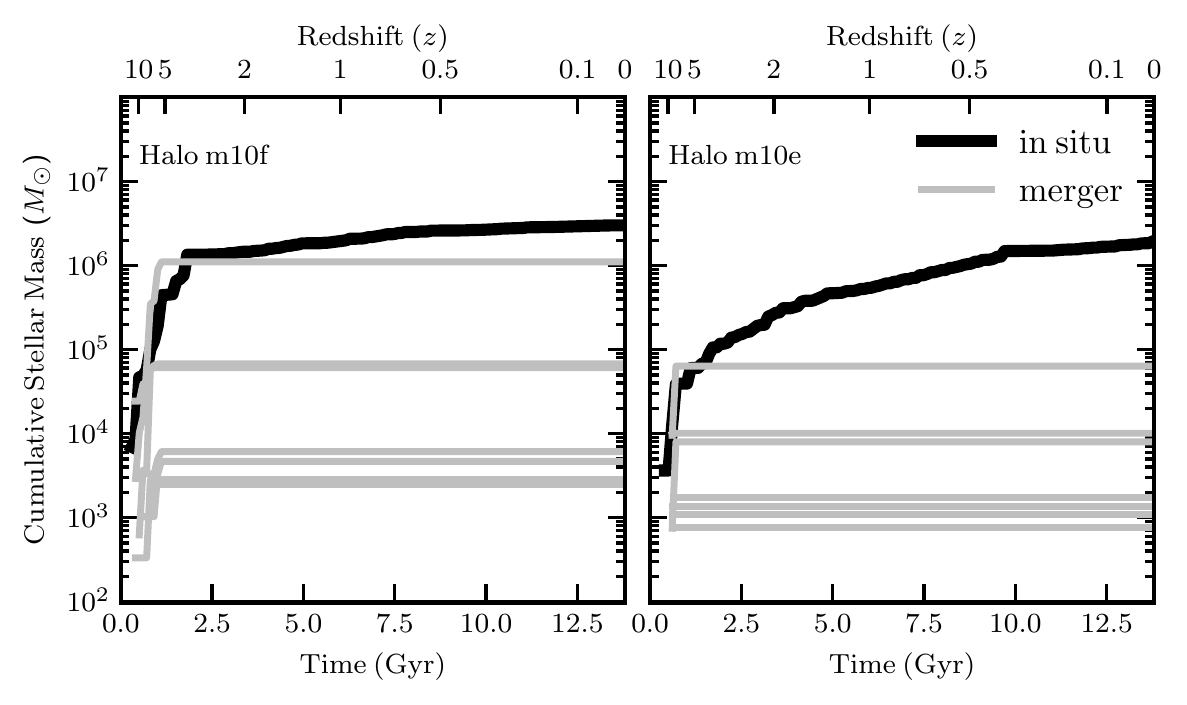}
  \caption{\textit{Left}: "Archaeological" stellar mass assembly history for halo m10f, measured from the birth times of all of the stars in the galaxy at $z=0$ (mimicking SFHs derived from resolved star observations in the Local Group). The in situ SFH is plotted in black while each population of stars that originated from a merger event is plotted separately in grey. The galaxy participated in a nearly equal-mass merger at early times, after the majority of SF has also taken place, and hence an appreciable fraction of the stars at $z=0$ originate from galaxy mergers (a rarity in our sample). \textit{Right}: The same as the left panel expect for halo m10e. Although it also had a nearly equal-mass merger, this halo has has substantial late-time star formation. Hence, even though it suffers a major merger early on in its life, the imprint left by $z=0$ is but a small fraction of the total stellar mass.}
  \label{fig:arch_SFH}
\end{figure*} 
To provide a clearer distinction between the population of stars formed in situ and those delivered from galaxy mergers, we now look at the star formation history of each population. To provide a more appropriate comparison with SFHs derived from observed CMDs, Figure~\ref{fig:arch_SFH} displays the "archaeological" star formation history for halo m10e (right) and m10f (left) ; this is created by using all the stars present at $z=0$ to calculate when a given fraction of the present-day stars were formed. In situ star formation is represented by the black line while those stars that originated from separate galaxy mergers are plotted in grey. Though both halos in Fig.~\ref{fig:arch_SFH} participate in nearly equal-mass galaxy mergers early in their formation, halo m10f, the smaller of the two, has enough continuous star formation over the rest of the simulation to overwhelm its early merger contribution. Meanwhile, halo m10e, which has formed a larger fraction of its $z=0$ stellar mass by the time it suffers its major merger, has an appreciable fraction of stars contributed by mergers in its stellar population at $z=0$. The majority of externally-produced stars are delivered from galaxy mergers that occur early in the formation of dwarfs. Several dwarfs have as much as 80$\%$ of their stellar mass from outside sources at early times. However after this early assembly phase, continuous in situ star formation results in all but two of the dwarfs having formed $>93\%$ of their stellar mass within the main progenitor by $z=0$. 

Each merger progenitor contributes a uniformly ancient stellar population to the main dwarf. This is consistent with observations showing that ultra-faint dwarfs contain exclusively ancient stellar populations (e.g., \citealt{Brown:2014,Weisz:2014a}). It is generally believed that since the virial temperature of ultra-faints' hosts is below the temperature of the photoionized IGM ($\sim 2\times 10^4$ K;  \citealt{Haiman:2003,Faucher-Giguere:2009}), gas accretion becomes highly inefficient after $z\sim5$. The gravitational potential of these tiny halos can barely retain gas that is heated by the UV background, and most of the remaining gas is not self-shielding. The median cooling time of CGM gas in these halos is only a few Gyr, so much of the gas would likely cool into the galaxy in the absence of external energy sources (e.g., \citealt{White:1991,Birnboim:2003,Keres:2005,Fielding:2017}). However, heating from the UV background and star formation, as well as frequent periodic out-flows, prevents CGM gas from reaching the central galaxy. Given the negligible quantity of gas delivered from both halo and galaxy mergers, we are able to obtain a `clean' representation of the main progenitor SFH of each dwarf uncontaminated by any outside influence by focusing only on stars born \textit{after} $z\sim3$. 

\subsection{Satellites}
Utilizing AHF, we are able create a subhalo catalogue for each of the dwarfs within our suite of simulations. By focusing only on subhalos that contain bound star particles and are beyond the central galaxy ($r>0.1R_\mathrm{vir}$), we are able to isolate the presence of any luminous satellites. Of the 15 dwarfs we have simulated, only two contain visible companions at a redshift of zero.  To test the convergence of these results, we have simulated all dwarfs at a resolution 8x lower than our fiducial resolution and have resimulated one of our paired dwarfs, halo m10b, at a resolution 8x higher as well. While halo m10e does have a luminous companion in both resolution levels, we find that m10b's companion is (and always has been) completely dark in the lowest resolution. 

\begin{figure*}
\includegraphics[width=\textwidth]{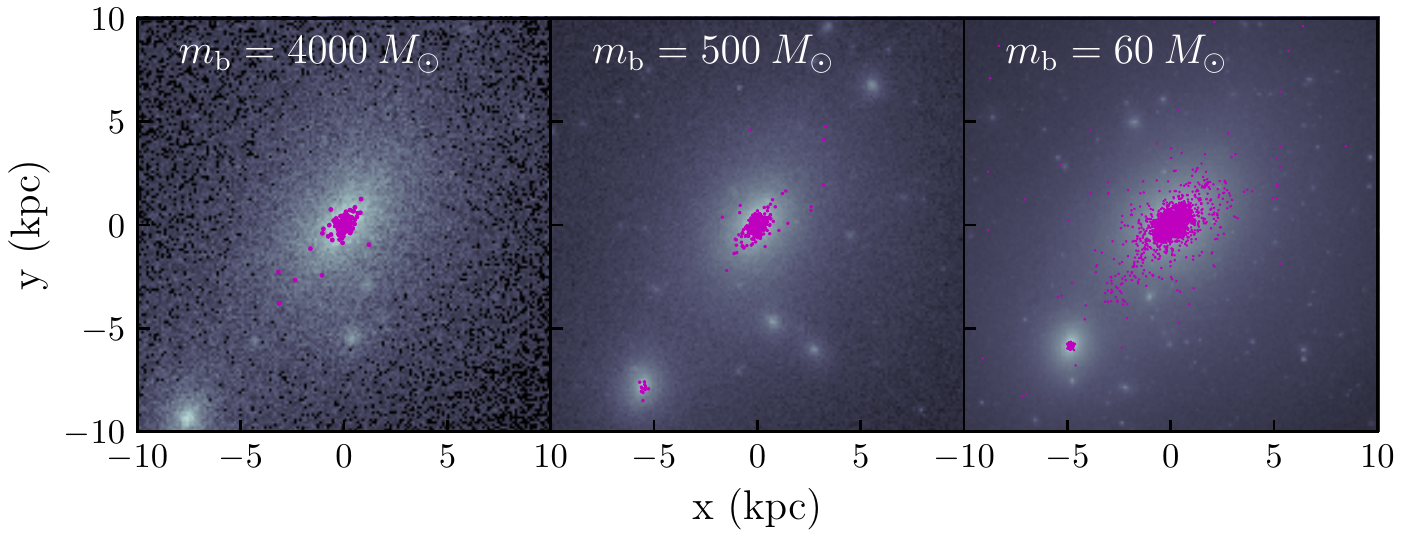}
\caption{\textit{Left}: 2D projected density map (gray-scale) of the dark matter for dwarf m10b$\_$low (halo m10b at our lower resolution) at $z=0$. Stars are plotted on top of the density map in magenta. m10b$\_$low's completely dark satellite companion can be seen in the lower left corner. \textit{Middle}: Same as left plot except at our fiducial resolution. We can see that the companion now has a population of bound stars. \textit{Right}: Same as previous two plots except for m10b$\_$high (m10b at our higher resolution); the satellite companion again is apparent in both dark matter and stars. Though all three versions of m10b have a prominent dark matter satellite, the companion is only luminous in the fiducial and high resolution versions.}
  \label{fig:2dsat}
\end{figure*}

In Fig. \ref{fig:2dsat}, we show a 2D projected density map of the dark matter distribution for each version of halo m10b (with resolution increasing from left to right). Plotted over the dark matter distribution are the star particles present in each halo (magenta). Halo m10b's companion can be seen in the lower left hand corner, with visibly bound star particles clustering within the satellite in the fiducial and high resolution versions. Figure \ref{fig:2dsat} hints that higher resolution simulations may be necessary to properly resolve 1:100 satellites around M$_\star \sim 10^6\,\msun$ dwarfs and will be absolutely necessary to study the existence of possible 1:1000 satellites.

Table~\ref{table:params} includes some basic properties of the all the satellites in our simulations. One particularly interesting point is that each satellite is surrounded by between $10^3$ and $10^4\,\msun$ of gas. However, this gas is warm/hot ($>10^5$~K) in every version of each satellite, meaning there is no fuel for star formation (the cooling times for this low-density $10^5$~K gas are very long). Indeed, star formation in these satellites has long been dormant, as indicated by the right panel in Fig.~\ref{fig:arch_SFH} for halo m10e's satellite. These minor galaxy mergers therefore bring \emph{only} ancient stellar populations with them. Focusing on the stellar half-mass radii radii, we notice a trend towards smaller sizes with increasing resolution. This is at least partially a resolution effect, as both m10b + m10e$\_$low's satellites have very few star particles and thus lend themselves to counting errors (discussed more in appendix \ref{sec:appendixa}). 

Looking at both Fig. \ref{fig:mvirs_v_t} and table \ref{table:params}, we see that each surviving satellite is accreted onto its main dwarf at relatively late cosmological time. In the case of halo m10e, its companion is on its first approach, having entered the virial radius at $z=0.36$. Meanwhile halo m10b's satellite has already had a flyby at $z\sim0.5$, briefly left the virial radius at $z\sim0.2$, and at $z=0.038$ began what is likely its final approach. The late-time accretion for the majority of present-day subhalos has been seen previously in dissipationless simulations \citep{Gao:2004}; in the context of dwarf galaxies, the scarcity of surviving satellites owes to the early assembly epoch of a typical dwarf. 

\section{Discussion and Conclusions}
\label{sec:discussion}

In an effort to better understand the origin of dwarf galaxy stellar populations, we have explored the merger histories of 15 cosmological zoom-in simulations of field dwarf galaxies in $10^{10}\,\msun$ halos from the \fire~project. Our simulation suite presents a clear picture of isolated dwarf galaxy formation with only minor contributions from baryonic material brought in via mergers: 13 of 15 galaxies form $>93\%$ of their stars in situ, and all 15 galaxies have at least 70\% of their stars formed in the main progenitor. The ultra-faint dwarfs that merge with our main galaxies or remain as satellites at $z=0$ are also driven by in situ processes: they all are  composed of uniformly ancient stars resulting from an initial burst of star formation, are unable to accrete new gas to form stars, and merge only with starless dark matter halos. 

\input{table2.tex}
Other research presents a different picture, with ultra-faint dwarf (UFD) stellar populations built up from star formation episodes in separate halos \citep{Simpson:2013,Jeon:2017}. One important distinction, however, is that the UV background in these simulations is turned on at a much later redshift. This is likely why they have luminous progenitors in halos nearly an order of magnitude smaller than in our simulations, as the timing of reionization has a major impact on the final stellar content of a dark matter halo by interrupting the cooling of gas onto lower-mass progenitors \citep{Simpson:2013}. Simpson et al.~also present an `early UV' version of their dwarf, with the UV background instead turning on at $z\sim11$, which bears a better resemblance to both our UV background implementation and our UFDs' resulting monolithic stellar populations. The properties of these UFDs are therefore highly sensitive to the timing of reionization; accordingly, UFDs can serve as a useful tool for learning about the reionization process.

\begin{figure}
\includegraphics[width=\columnwidth]{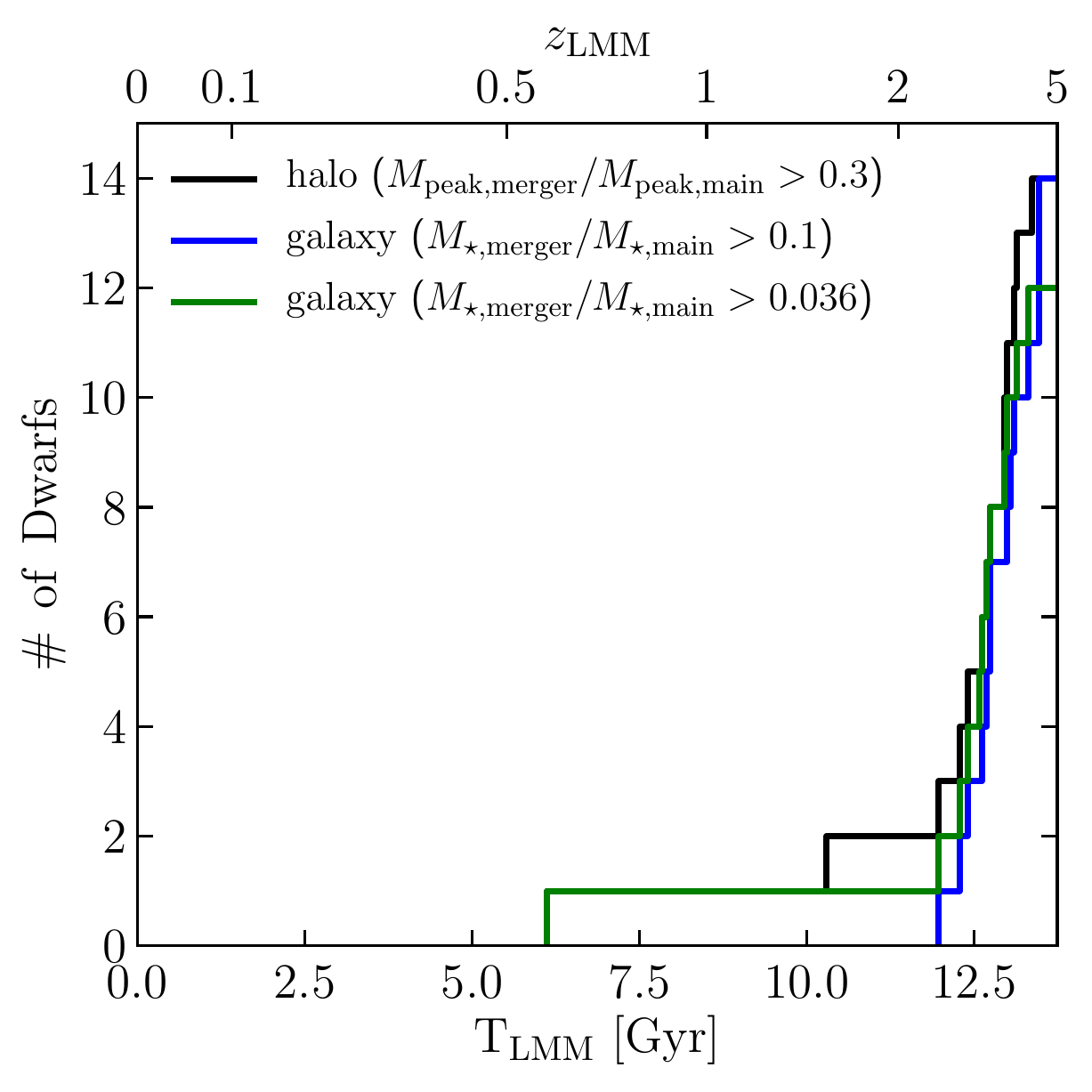}
  \caption{Cumulative distribution of lookback time to the last major merger, T$_{\rm LMM}$, calculated three different ways. The black line shows the distribution of the last major halo merger (defined as M$_{\rm peak,merger}$/M$_{\rm peak,main}>0.3$) while the blue line corresponds to the last major galaxy merger (defined as M$_{\rm \star,merger}$/M$_{\rm \star,main}>0.1$). If we use the M$_\star$-M$_{\rm peak}$ relation derived from our simulation suite, we find that M$_{\rm peak,merger}$/M$_{\rm peak,main}>0.3$ corresponds to M$_\star$-M$_{\rm peak}>0.036$; the resulting distribution of merger times satisfying this criterion is plotted as a green line. This distribution provides a good match to the distribution of last major halo merger times, indicating that a carefully calibrated stellar mass-halo mass relation applied to DMO simulations can reproduce the distribution of merger times found in hydrodynamic simulations.}
  \label{fig:zlmm}
\end{figure} 
Galaxy mergers have been invoked to explain both the presence of a metallicity gradient in  dwarfs \citep{Benitez-Llambay:2016} and the rekindling of star formation in `two-component' dwarfs \citep{Benitez-Llambay:2015}. The former of these effects has already been found with m10q in \cite{El-Badry:2016} while the latter can be directly seen in halo m10b: around $z\sim1$, the main progenitor experiences a galaxy merger (see Fig.~\ref{fig:mvirs_v_t}) that spurs new star formation.  Halo mergers have also been shown to give a strong rise in the star formation rate of a dwarf \citep{Starkenburg:2016a}. To have a significant impact, a dark satellite must have at least 10$\%$ of the mass of the host. We witness such an interaction in our suite:  at $z\sim2$, the main progenitor of halo m10v merges with 5 dark halos, each with a merger ratio > $1:10$. This interaction is synchronized with a compression of the gas in the main halo and a subsequent up-tick in star formation.

Previous studies have relied on dissipationless simulations to make statements on the galaxy formation of dwarfs (though see \citealt{Munshi:2017}). \cite{Deason:2014} studied the frequency of dwarf mergers using the dissipationless ELVIS simulations \citep{Garrison-Kimmel:2014}. They assigned a stellar mass to each halo by using a modified stellar mass to (sub)halo dark matter mass relation from \cite{Behroozi:2013b}. Their results are comparable to the black line in Fig. \ref{fig:zlmm}, which depicts the halos in our simulation that satisfy their criteria for a halo major merger. Using this framework, we predict that one of our halos would have experienced a major merger since $z=1$. However, if we instead use the actual galaxy major mergers that occur in our simulations, we conclude that no galaxy major merger has occurred in over 12 Gyr. Since \citet{Deason:2014}'s definition of a galaxy major merger is dependent on the specific modified stellar mass to (sub)halo dark matter mass relation they utilize, we also derive our own major galaxy merger criteria using the stellar mass to (sub)halo dark matter mass relation from our simulation suite. Using this criteria, we obtain a result (green line) that matches the halo major merger prediction (black) quite well. Fig.~\ref{fig:zlmm} demonstrates that while DMO simulations are accurate at determining the timing of major mergers in the majority of dwarfs, a statistical sample of hydrodynamical dwarf simulations will be necessary to constrain the exact fraction of dwarf galaxies with major mergers after early comic times.

Dissipationless simulations have also been used to try to infer the number of ultra-faint satellites we should expect around dwarf galaxies at the present day. \citet{Sales:2013} utilized abundance matching by identifying primary satellite systems in galaxy catalogues constructed from the SDSS and comparing them with predictions from a semianalytic mock galaxy catalogue based on the Millennium-II Simulation \citep{Boylan-Kolchin:2009}. They found a 1:1000 satellite for each dwarf and a 1:100 satellite for every 3-5 dwarfs when considering central galaxies with stellar masses below $\sim10^{10}\,\msun$. However, this analysis was constrained to galaxies with M$_\star>10^{6}\,\msun$. According to the abundance matching of \citet{Wheeler:2015a}, isolated halos with $\mvir\sim10^{10}\,\msun$ will have one or more subhalos that could host $\mstar$ >3000 $\msun$ satellites about 35$\%$ of the time. Our simulations only yield 2 roughly 1:100 satellites for our suite of 15, which is slightly fewer than what the dissipationless simulations predict. However, as is clear from the top panel in the right plot of Fig.~\ref{fig:mstar_mhalo}, the transition from dark halos to UFDs lacks a sharp boundary. If we take into account that $\sim80\%$ of halos at this mass are dark in our simulations, the dissipationless simulations arrive at a fairly similar prediction to ours. Extrapolating the number and timing of galaxy major mergers from the halo merger history alone provides a similar prediction to the galaxy merger histories in our simulation; however, this extrapolation has significant uncertainty as we continue to push to dimmer and dimmer galaxies.

Our hydrodynamical simulations, with a self-consistent treatment of feedback, provide a detailed view of galaxy assembly in isolated field dwarfs that does not rely on extrapolating from results of dissipationless simulations. We find that isolated dwarf galaxies assemble in a relatively insular manner: on average, they experience only $\sim$5 galaxy mergers throughout their lifetime, and the vast majority of such mergers contribute negligibly in terms of stellar mass. The stellar population of an isolated dwarf galaxy observed at $z=0$ is predominantly a reflection of the main progenitor's star formation itself as opposed to a patchwork of many different galaxies' star formation. The few galaxy mergers occur early in the assembly of the simulated dwarfs, before redshift 4 and possibly around the time of reionization. This is roughly in line with what is predicted by interpreting DMO simulations. Also of note is that two of our dwarf galaxies have luminous satellite companions at $z=0$. Though baryonic feedback has strong effects on galaxy formation at this mass scale, our simulations still predict the presence of satellites around isolated field dwarfs. 

Observations of the near field will serve as a crucial tool in the upcoming \textit{JWST} era.  Archaeological studies of resolved stellar populations in the Local Group will not only probe regions larger than the HUDF and any deep \textit{JWST} fields but also have the promise to probe  a cosmologically representative region for halos with $M_\mathrm{vir}(z=7)\lesssim2\times10^9\,\msun$ \citep{Boylan-Kolchin:2016}. An unbiased view of faint galaxy populations at early times will have strong implications for UFDs' role in reionization. For example, a combination of the stellar fossil record of these UFDs in the Local Group with population synthesis modeling may be able to probe the faint end of the high-$z$ UV luminosity function and reveal a possible turn-over in the luminosity function \citep{Boylan-Kolchin:2015,Weisz:2017}. Perhaps our most important result in this context is that the stellar populations of $z=0$ field dwarfs should mainly reflect in situ star formation in one main progenitor as opposed to the hierarchical assembly of many ancestors.

\section*{Acknowledgments}
MBK and AF acknowledge support from the National Science Foundation (grant AST-1517226). MBK was also partially supported by NASA through HST grants AR-12836, AR-13888, AR-13896, AR-14282, AR-14554, GO-12914, and GO-14191 awarded by the Space Telescope Science Institute (STScI), which is operated by the Association of Universities for Research in Astronomy (AURA), Inc., under NASA contract NAS5-26555. JSB was supported by  NSF AST-1518291, HST-AR-14282, and HST-AR-13888. Support for PFH was provided by an Alfred P. Sloan Research Fellowship, NASA ATP Grant NNX14AH35G, and NSF Collaborative Research Grant \#1411920 and CAREER grant \#1455342. CAFG was supported by NSF through grants AST-1412836, AST-1517491, AST-1715216, and CAREER award AST-1652522, and by NASA through grant NNX15AB22G. DRW is supported by a fellowship from the Alfred P. Sloan Foundation. EQ was supported by NASA ATP grant 12-APT12-0183, a Simons Investigator award from the Simons Foundation, and the David and Lucile Packard Foundation. DK was supported by NSF grant AST-1715101 and the Cottrell Scholar Award from the Research Corporation for Science Advancement. AW was supported by a Caltech-Carnegie Fellowship, in part through the Moore Center for Theoretical Cosmology and Physics at Caltech, and by NASA through grant HST-GO-14734 from STScI. This work used computational resources of the University of Texas at Austin and the Texas Advanced Computing Center (TACC; \url{http://www.tacc.utexas.edu}), the NASA Advanced Supercomputing (NAS) Division and the NASA Center for Climate Simulation (NCCS) through allocations SMD-15-5902, SMD-15-5904, SMD-16-7043, and SMD-16-6991, and the Extreme Science and Engineering Discovery Environment (XSEDE, via allocations TG-AST110035, TG-AST130039, and TG-AST140080), which is supported by National Science Foundation grant number OCI-1053575. 
\label{lastpage}

\bibliography{bibliography}
\appendix
\section{Resolution and Convergence}
\label{sec:appendixa}
\begin{figure*}
\includegraphics[width=0.995\textwidth]{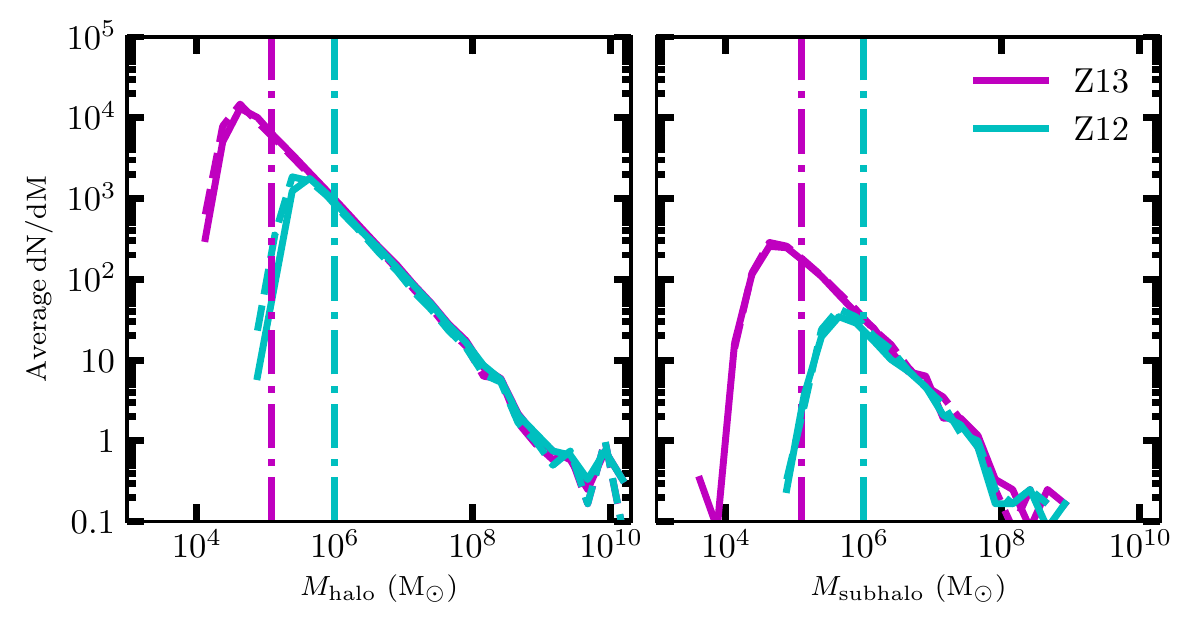}
  \caption{\textit{Left}: The halo mass function, averaged over all of the halos in our 15 simulations that have at least 95$\%$ high resolution particles. The magenta lines denote our fiducial resolution, Z13, while the cyan lines correspond to our low resolution, Z12 (8 times poorer mass and 2 times poorer force resolution). The solid lines are computed from the hydrodynamical simulations, while the dashed lines (which are difficult to see, as they are covered by the solid lines) are computed from DMO simulations. The vertical dashed-dot lines mark the 'resolved' threshold for each resolution level. \textit{Right}: Identical to the left plot but instead displays the average mass function for subhalos only. The two panels display excellent agreement for resolved dark matter structures, both between dissipationless and hydrodynamical runs as well as between resolution levels.}
  \label{fig:halo_func}
\end{figure*}

\begin{figure}
\includegraphics[width=\columnwidth]{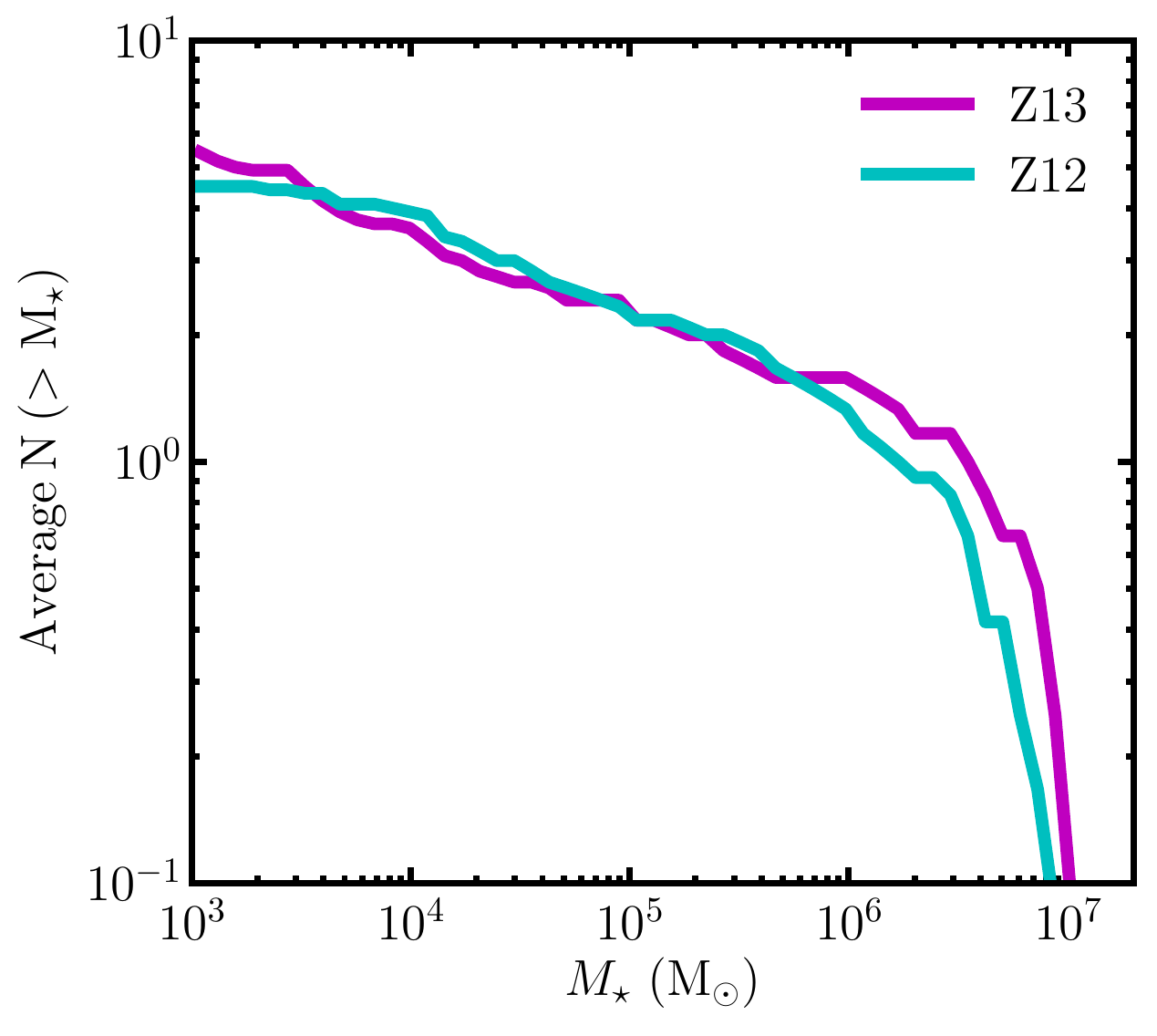}
  \caption{Average cumulative stellar mass function over all of the galaxies in our 15 simulations that have at least 95$\%$ high resolution particles. The magenta lines denote our fiducial resolution, Z13, while the cyan lines mark our low resolution, Z12. Overall, there is excellent convergence in the average stellar mass function.}
  \label{fig:galaxy_func}
\end{figure}

Figure \ref{fig:halo_func} examines the convergence of the average halo and subhalo mass functions for the dwarfs in our suite at two resolution levels (separated by a factor of 8 in mass between our lowest resolution -- Z12, in cyan -- and our highest resolution, Z13, which is plotted in magenta) at $z=0$. DMO runs are plotted as dotted lines, while hydrodynamic runs are shown as solid lines. Dot-dashed lines mark the `resolved' threshold for each resolution level. Overall, the agreement between DMO and hydro is excellent in both plots. Similarly, there is nearly perfect agreement between low and high resolution runs above the `resolved' threshold of $\sim15$ particles (M$_\mathrm{vir}\sim3\times10^5\,\msun$) for the low-resolution runs. Though the high-resolution runs are able to resolve halos that are nearly ten times lower in mass relative to the low-resolution runs, this does little to affect the `resolved' galaxy stellar mass function in our simulations.  Plotting the average cumulative stellar mass function for our runs in fig. \ref{fig:galaxy_func}, we find agreement between the two resolution levels down to the resolution limit in the Z13 runs. This convergence in stellar mass is a result of most of these galaxies residing in well-resolved $\sim10^8\,\msun$ halos, as Figure \ref{fig:mstar_mhalo} illustrates. 

\begin{figure}
\includegraphics[width=\columnwidth]{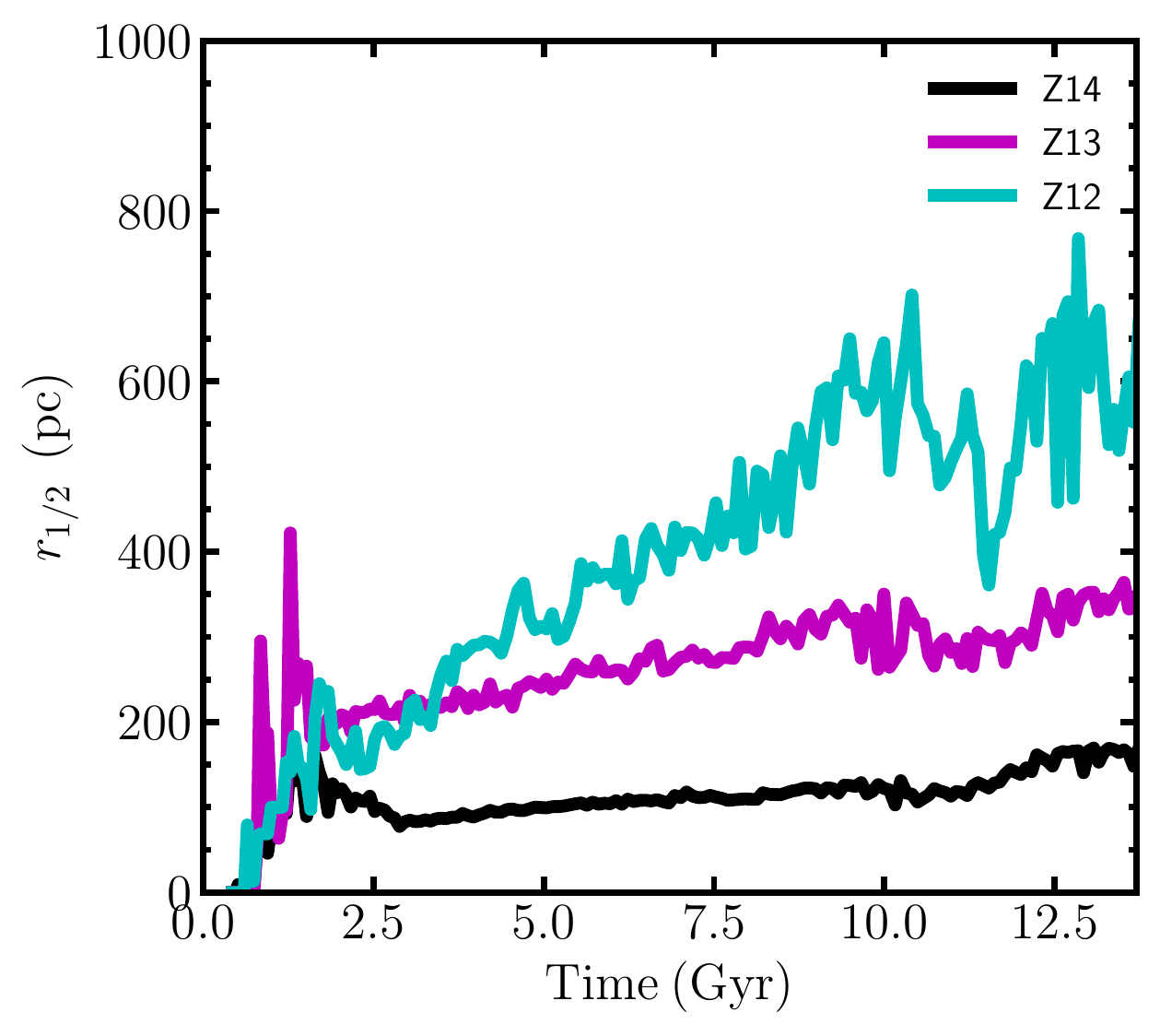}
  \caption{The stellar half-mass radius for halo m10b across cosmic time across three different levels of resolution. The lines are colored similarly to Figures~\ref{fig:halo_func}-\ref{fig:galaxy_func}, with the addition of the black line to represent our ultra-high resolution run (Z14). At increasingly high resolution, the stellar half-mass radius becomes smaller and is more stable over time. }
  \label{fig:halflight}
\end{figure}

Though there is excellent convergence in the mass functions of our simulations, increasing the resolution does lead to a decreasing stellar half-mass radius ($r_{1/2}$) for halo m10b. Figure \ref{fig:halflight} explores this further by plotting $r_{1/2}(t)$ for halo m10b at all three resolution levels. Though all versions of halo m10b start out with roughly the same physical size, the lowest resolution version nearly triples in size over time. The more better-resolved runs maintain their physical size throughout the entirety of the simulation. Given the bursty nature of star formation in these simulations, lower-resolution versions -- which do not resolve the central dark matter potential as well -- may see excess "heating" of stars, leading to larger sizes \citep{El-Badry:2017}. We will explore this issue further in a future paper.

\end{document}

%% file: table1.tex
\begin{table*}
  \caption{\textit{Global properties at $z=0$ for three different resolution levels of halo m10b.}. Columns:
(1) Baryon particle mass;  
(2) Virial mass; 
(3) Maximum amplitude of rotation curve; 
(4) Stellar mass of the central galaxy [defined as $\mstar(<0.1\,\rvir)$];
(5) Mass of gas below $T=10^4$ K within $\rvir$; 
(6) Total baryon fraction within $\rvir$, scaled to cosmic baryon fraction $f_{\rm b}$; 
(7) 3D stellar half-mass radius;
(8) Ratio of total mass to stellar mass within the stellar half-mass radius; 
(9) Ratio of virial mass in hydrodynamic run to virial mass in DMO run (after correcting the DMO virial mass for $f_{\rm b}$).
}
 \begin{threeparttable}
\centering 
\begin{tabular}{lcccccccccc}
\hline
\hline  
& $m_\mathrm{bary} $&$M_{\mathrm{vir}}$ &$\vmax$& $\mstar$ & $\mathrm{M_{gas,cold}}$ &$f_\mathrm{baryon}/f_\mathrm{b}$& $\rh$&$\mathrm{M_{dyn}}/\mstar$ & $M_\mathrm{hydro}/M_\mathrm{dmo}$\\ 
  &  $[\msun]$&$[\msun]$ &[$\kms$] & $[\msun]$ & $[\msun]$ & --&[pc]&$(<\rh)$ & --\\
\hline 
Halo & (1) & (2) & (3) & (4) & (5) & (6) & (7) & (8) & (9)\\
\hline 
m10b$\_$low  &$4000$&$9.32\times10^9$& 31.61 & $4.13\times10^5$ & $ 3.45\times10^{5}$& 0.107 & $556$& 209.77  & 0.957 \\
m10b  &$500$&$9.29\times10^9$& 31.51 & $4.65\times10^5$ & $ 6.63\times10^{6}$& 0.113 & $340$& 96.56  & 0.962 \\
m10b$\_$high  &$62.5$&$9.22\times10^9$& 32.02 & $8.59\times10^5$ & $ 1.34\times10^{6}$& 0.113 & $260$& 26.12  & 0.961\\
\hline  
\end{tabular}
 \end{threeparttable}
 \label{table:cprops}
\end{table*}

%% file: table2.tex
\begin{table*}
  \caption{\textit{Global properties at $z=0\,$ for satellites of simulated field galaxies with $\mvir \approx 10^{10} \,\msun$}. Columns:
(1) Virial mass of satellite; 
(2) Maximum amplitude of rotation curve; 
(3) Stellar mass of the satellite galaxy;
(4) Mass of gas ; 
(5) Redshift of initial accretion onto main dwarf;
(6) 3D stellar half-mass radius;
(7) Satellite distance to host;
(8) Ratio of virial mass of satellite to virial mass of host galaxy; 
(9) Ratio of stellar mass of satellite to stellar mass of host galaxy.
\label{table:params}
}
 \begin{threeparttable}
\centering 
\begin{tabular}{lcccccccccc}
\hline
\hline  
& $M_{\mathrm{vir}}$ &$\vmax$& $\mstar$ & $\mathrm{M_{gas}}$ &$z_\mathrm{acc}$& $\rh$& $\mathrm{R_{to\:host}}$&$\mathrm{M_{vir,sat}/M_{vir,host}}$&$\mathrm{M_{\star,sat}/M_{\star,host}} $\\ 
  &  $[\msun]$ &[$\kms$] & $[\msun]$ & $[\msun]$ & --&[pc]&[kpc]& --&--\\
\hline 
Host Halo & (1) & (2) & (3) & (4) & (5) & (6) & (7) & (8) & (9)  \\
\hline 
m10b$\_$low  &$6.81\times10^8$& 18.58 & $0$ & $ 7.89\times10^3$& 0.485 & -- & 33.672& 0.073 & 0\\
m10b  &$7.38\times10^8$& 18.72 & $4.24\times10^3$ & $ 1.50\times10^3$& 0.485 & $412$  & 32.499& 0.079 & 0.009 \\
m10b$\_$high  &$6.63\times10^8$& 18.92 & $1.02\times10^4$ & $ 2.25\times10^3$& 0.485 & $148$ & 26.981& 0.071 & 0.022 \\
\\
m10e$\_$low &$\;4.80\times10^{8}$& 15.66& $5.98\times10^3$ & $ 3.99\times10^{3}$& 0.367 & $783$ &32.621& 0.047 & 0.011  \\
m10e &$\;4.78\times10^{8}$& 15.85& $1.36\times10^4$ & $ 1.00\times10^{3}$& 0.367 & $311$ &36.508& 0.047 & 0.007  \\
\hline  
\end{tabular}
 \end{threeparttable}
\end{table*}